\documentclass[conference,letterpaper]{IEEEtran}

\usepackage[T1]{fontenc}
\usepackage{graphicx}
\usepackage{cite}
\usepackage{subcaption}

\usepackage{overpic}
\usepackage{amssymb}
\usepackage{amsmath}
\usepackage{amsthm}
\usepackage{array}
\usepackage{color}
\usepackage{url}

\IEEEoverridecommandlockouts

\theoremstyle{definition}

\newcommand{\vect}[1]{\boldsymbol{#1}}

\def\kron{\otimes}

\def\Htran{\mbox{\tiny $\mathrm{H}$}}
\def\Ttran{\mbox{\tiny $\mathrm{T}$}}
\def\CN{\mathcal{N}_{\mathbb{C}}} 
\def\imagunit{\mathsf{j}} 
\def\sinc{\mathrm{sinc}}

\begin{document}

\title{Optimizing a Binary Intelligent Reflecting Surface for OFDM Communications under Mutual Coupling}

\author{
\IEEEauthorblockN{Emil Bj{\"o}rnson
\thanks{This work was supported by the FFL18-0277 grant from the Swedish Foundation for Strategic Research.}}
\IEEEauthorblockA{Department of Electrical Engineering, Link\"{o}ping University, Link\"{o}ping, Sweden}
\IEEEauthorblockA{Department of Computer Science, KTH Royal Institute of Technology, Kista, Sweden (emilbjo@kth.se)}
}

\maketitle

\begin{abstract}
An intelligent reflecting surface (IRS) can greatly improve the channel quality over a frequency-flat channel, if it is configured to reflect the incident signal as a beam towards the receiver.
However, the fundamental limitations of the IRS technology become apparent over practical frequency-selective channels, where the same configuration must be used over the entire bandwidth.
In this paper, we consider a wideband orthogonal frequency-division multiplexing (OFDM) system that is supported by a fairly realistic IRS setup with two unbalanced states per element and also mutual coupling. 
We describe the simulation setup considered in the IEEE Signal Processing Cup 2021, propose a low-complexity solution for channel estimation and IRS configuration, and evaluate it on that setup.
\end{abstract}

\IEEEpeerreviewmaketitle

\section{Introduction}

The wireless propagation environment is conventionally uncontrollable but this can be changed using an intelligent reflecting surface (IRS) \cite{Wu2018a}, also known as a reconfigurable intelligent surface \cite{Huang2018a}. An IRS consists of a two-dimensional array of passive elements that will (diffusely) reflect incident radio waves. In contrast to conventional surfaces with uniform properties, each IRS element has a configurable impedance that can be tuned to modify the amplitude and phase-shift of the reflected signal. 
The non-uniform phase-shift pattern over the IRS determines the directivity of the reflected signal; for example, the signal from an access point (AP) can be reflected in the shape of a beam towards the intended receiver \cite{RIS_SPMAG}.

The IRS configuration problem is well-studied for frequency-flat channels \cite{Wu2018a,Huang2018a,Abeywickrama2020a,RIS_SPMAG,Renzo2020b}, however, most practical systems operate over frequency-selective channels.
An IRS is less effective over such channels because it is can only take a single configuration at a time, while different subcarriers in orthogonal frequency-division multiplexing (OFDM) systems prefer different configurations.
The joint channel estimation and IRS configuration for OFDM systems are studied in \cite{Zheng2020,Yang2020a,Lin2020a}, where heuristic algorithms with varying complexity are proposed.
A key assumption in these prior works is that the IRS elements can be perfectly tuned, in the sense of having a constant amplitude, continuous phase configuration, and no mutual coupling. A practical IRS might not satisfy any of these ideal conditions \cite{Abeywickrama2020a,RIS_SPMAG}, which calls for new solutions.

In this paper, we will go beyond the state-of-the-art by considering an OFDM system with only two unbalanced states per IRS element and unknown coupling between the adjacent elements. We propose new algorithms for channel estimation and configuration, and analyze them using the dataset from the IEEE Signal Processing Cup 2021 \cite{SP_CUP2021}. The code and dataset are available at \url{https://github.com/emilbjornson/SP_Cup_2021}

\section{System Model}

This paper considers a setup where a single-antenna AP communicates with a multitude of user equipments (UEs), with the support of an IRS consisting of $N$ controllable elements. One UE is served at a time over a wideband channel with a bandwidth $B$. Hence, we can consider the UEs individually but must design an algorithm for IRS configuration that enables efficient switching between the UEs.
We assume  the transmission is carried out using OFDM with a unit-energy sinc-function as the pulse-shape filter, as in the model derived in  \cite{RIS_SPMAG}.
Let $\{ x[k] \}$ denote the transmitted discrete-time signal in the complex baseband domain. The corresponding received discrete-time signal sequence $\{ z[k] \}$ is given by
\begin{equation}
z[k] =  \sum_{\ell=0}^{M-1} h_{\vect{\theta}}[\ell] x[k-\ell] + w[k]
\end{equation}
where $\{ h_{\vect{\theta}}[\ell] : \ell=0,\ldots,M-1 \}$ is the finite impulse response (FIR) filter that describes the wideband  channel in the time-domain and  $\{ w[k] \}$ is the receiver noise. The FIR filter has $M$ taps, which depend on the physical propagation paths as well as the IRS configuration, where the latter is denoted by $\vect{\theta}$.
Tap $\ell$ is given as
\begin{equation}
h_{\vect{\theta}}[\ell]  = h_{d}[\ell] + \vect{v}_{\ell}^{\Ttran} \vect{\omega}_{\vect{\theta}}
\end{equation}
where $h_{d}[\ell]$ is the uncontrollable channel (containing all multipath components that are not involving the IRS), $\vect{v}_{\ell} \in \mathbb{C}^{N}$ is the cascaded channels via each of the $N$ elements (for the hypothetical case where each element reflects the entire incident signal without delay), and $\vect{\omega}_{\vect{\theta}} \in \mathbb{C}^N$ contains the actual reflection coefficients of the IRS that determine the amplitude losses and phase shifts (it is the same for all $\ell$).

We consider communication over a channel bandwidth of $B$\,Hz, which is also the symbol rate.
The OFDM transmission makes use of an $M-1$ length cyclic prefix and creates $K>M$ subcarriers. Hence, a block of $K+M-1$ time-domain signals is transmitted to create one OFDM block with $K$ parallel subcarriers using the discrete Fourier transform (DFT):
\begin{equation} \label{eq:system-model}
\bar{z}[\nu] = \bar{h}_{\vect{\theta}}[\nu] \bar{x}[\nu] + \bar{w}[\nu], \quad \nu = 0, \ldots, K-1
\end{equation}
where the DFTs are $\bar{z}[\nu] = \frac{1}{\sqrt{K}} \sum_{k=0}^{K-1} z[k] e^{-\imagunit 2 \pi k \nu /K}$, $\bar{x}[\nu] = \frac{1}{\sqrt{K}} \sum_{k=0}^{K-1} x[k] e^{-\imagunit 2 \pi k \nu /K}$, $\bar{h}_{\vect{\theta}}[\nu] = \sum_{k=0}^{M-1} h_{\vect{\theta}}[k] e^{-\imagunit 2 \pi k \nu /K}$, and $\bar{w}[\nu] = \frac{1}{\sqrt{K}} \sum_{k=0}^{K-1} w[k] e^{-\imagunit 2 \pi k \nu /K}$.\footnote{Notice that the normalized DFTs must be taken for the signals and noise but not for the channel response to obtain a valid pseudo-baseband model \cite{RIS_SPMAG}.}

If we let $\odot$ denote the Hadamard (element-wise) product, we can write the system model in \eqref{eq:system-model} in vector form as:
\begin{equation}
\underbrace{\begin{bmatrix} \label{eq:system-model2}
\bar{z}[0]  \\
\vdots \\
\bar{z}[K-1] 
\end{bmatrix}}_{=\vect{\bar{z}}} = \underbrace{\begin{bmatrix}
\bar{h}_{\vect{\theta}}[0]  \\
\vdots \\
\bar{h}_{\vect{\theta}}[K-1] 
\end{bmatrix}}_{=\vect{\bar{h}}_{\vect{\theta}}} \odot \underbrace{\begin{bmatrix}
\bar{x}[0]  \\
\vdots \\
\bar{x}[K-1] 
\end{bmatrix}}_{=\vect{\bar{x}}} + \underbrace{\begin{bmatrix}
\bar{w}[0]  \\
\vdots \\
\bar{w}[K-1] 
\end{bmatrix}}_{=\vect{\bar{w}}}.
\end{equation}
The input-output relation of one OFDM block can be written in short form as
\begin{equation} \label{system-model}
\vect{\bar{z}} = \vect{\bar{h}}_{\vect{\theta}} \odot \vect{\bar{x}} + \vect{\bar{w}},
\end{equation}
where all the vectors are of length $K$. We can also notice that 
\begin{equation} \label{eq:OFDM-response}
\bar{\vect{h}}_{\vect{\theta}}
= \vect{F} \begin{bmatrix}
h_{d}[0] + \vect{v}_{0}^{\Ttran} \vect{\omega}_{\vect{\theta}} \\
\vdots \\
h_{d}[M-1] + \vect{v}_{M-1}^{\Ttran} \vect{\omega}_{\vect{\theta}}
\end{bmatrix}
= \vect{F} \left( \vect{h}_d + \vect{V}^{\Ttran} \vect{\omega}_{\vect{\theta}} \right)
\end{equation}
where $\vect{h}_d = [ h_{d}[0] , \ldots, h_{d}[M-1]]^{\Ttran}$ gathers all the uncontrollable channel components, 
$\vect{V}= [\vect{v}_0, \ldots, \vect{v}_{M-1}]  \in \mathbb{C}^{N \times M}$ gathers all components containing the controllable propagation channels, and $\vect{F}$ is a $K \times M$ DFT matrix with the $(\nu,k)$th element being $e^{-\imagunit 2 \pi k \nu / K}$. 

Different configurations $\vect{\theta}$ result in different channel vectors $\vect{\bar{h}}_{\vect{\theta}}$.
For a given configuration, equal power allocation, and perfect channel knowledge at the receiver, the sum information rate over the subcarriers in \eqref{eq:system-model} is
\begin{align} \label{eq:rate-OFDM}
    R = \frac{B}{K+M-1} \sum_{\nu=0}^{K-1} \log_2 \left( 1 + \frac{P | \bar{h}_{\vect{\theta}}[\nu] |^2}{B N_0} \right) \,\, \textrm{bit/s}
\end{align} 
where $P$ is transmit power and $N_0$ is the noise power spectral density (i.e., $\bar{w}[k] \sim \CN(0,N_0)$).
The rate can be maximized with respect to $\vect{\theta}$. We will do that in a setup where each IRS element can only take one of two states and where mutual coupling also tampers with the controllability.

\section{Description of the Specific \\ IRS-Aided Communication Setup}

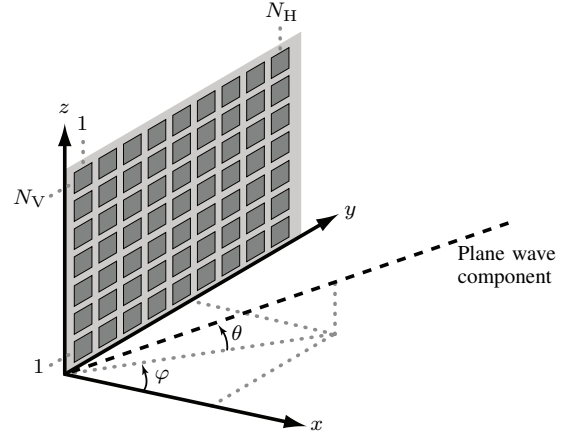
\begin{figure}[t!]
	\centering 
	\begin{overpic}[width=.85\columnwidth,tics=10]{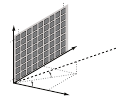}
		\put(54,1){\footnotesize $x$}
		\put(60,39){\footnotesize $y$}
		\put(9.5,57){\footnotesize $z$}
		\put(26.5,10){\footnotesize  $\varphi$}
		\put(40,17){\footnotesize  $\theta$}
		\put(5,11){\footnotesize  $1$}
		\put(1.5,41){\footnotesize  $N_{\mathrm{V}}$}
		\put(12.7,54){\footnotesize  $1$}
		\put(46,74){\footnotesize  $N_{\mathrm{H}}$}
		\put(80,31){\footnotesize  Plane wave}
		\put(80,27){\footnotesize  component}
\end{overpic} 
	\caption{The geometry of an IRS consisting of $N_{\mathrm{H}}$ elements per row and $N_{\mathrm{V}}$ elements per column.} \vspace{-3mm}
	\label{figure_geometric_setup}  
\end{figure}

We will now describe the IRS setup that will be analyzed in this paper and that has also been considered in the IEEE Signal Processing Cup, where a dataset was provided for this setup but without disclosing the details about the non-linear models that are described in this section. The code and dataset is available at  \cite{SP_CUP2021}.
The considered IRS consists of $N=4096$ entries in the form of a uniform planar array with $N_{\mathrm{H}}=64$ elements per horizontal row and $N_{\mathrm{V}}=64$ elements per vertical column. 
The array is illustrated in Fig.~\ref{figure_geometric_setup} and is deployed in the $yz$-plane according to the coordinate system defined in the figure.
The array is designed for communication at the carrier frequency $f_c=4$ GHz using a bandwidth of $B=10$ MHz.
The horizontal and vertical elements spacing are $0.4 \lambda$, where $\lambda=7.5$\,cm is the wavelength at the carrier frequency.
Each element has a directivity pattern of 
$G(\varphi,\theta) = \cos^2(\varphi) \cos(\theta)$, 
where $\varphi$ is the azimuth angle and $\theta$ is the elevation angle defined in Fig.~\ref{figure_geometric_setup}.

We assume the first element is located in the origin of the coordinate system and the elements
 are then indexed row-by-row by $n\in[1,N]$, thus the location of the $n$th element is
\begin{equation}
\vect{u}_n = [ 0, \, \,\, i(n) 0.4 \lambda,  \,\,\, j(n) 0.4 \lambda]^{\Ttran}
\end{equation}
where 
$i(n) =\mathrm{mod}(n-1,N_\mathrm{H})$ and
$j(n) =\left\lfloor(n-1)/N_\mathrm{H}\right\rfloor$
are the horizontal and vertical indices of element $n$, respectively, $\mathrm{mod}(\cdot,\cdot)$ denotes the modulus operation, and $\lfloor \cdot \rfloor$ truncates the argument.
If a plane wave is impinging on the IRS from the angle pair $(\varphi,\theta)$, the array response vector is \cite[Sec.~7.3]{massivemimobook}
\begin{equation}\label{array-response}
\vect{a}(\varphi,\theta) = \sqrt{G(\varphi,\theta)} \left[e^{\imagunit\vect{k}(\varphi,\theta)^{\Ttran}\vect{u}_1},\dots,e^{\imagunit\vect{k}(\varphi,\theta)^{\Ttran}\vect{u}_N}\right]^{\Ttran}
\end{equation}
where  
$\vect{k}(\varphi, \theta) = \frac{2\pi}{\lambda}\left[\cos(\theta) \cos(\varphi),  \cos(\theta) \sin(\varphi),  \sin(\theta)\right]^{\Ttran}$ is the wave vector. The array response vector determines the phase-shifts between IRS elements relative to element $1$.

\subsection{Reflection Modeling}

Element $n$ is modeled according to \cite{Abeywickrama2020a} as a lumped circuit with the impedance
\begin{equation*}
    Z_n(C_n,f) = \frac{\imagunit 2\pi f L_1 \left( \imagunit 2\pi f  L_2 + \frac{1}{\imagunit 2\pi f C_n} + R \right)}{\imagunit 2\pi f L_1+ \imagunit 2\pi f  L_2 + \frac{1}{\imagunit 2\pi f C_n} + R },
\end{equation*}
where $L_1=2.8$\,nH and $L_2 = 0.8$\,nH are inductances in different layers, and $R=1\,\Omega$ is the effective resistance. The 
effective capacitance $C_n$  can be controlled using a diode.
The reflection coefficient of this IRS element is computed as
\begin{equation} \label{eq:reflection-coefficient}
\Gamma(C_n,f) = \frac{Z_n(C_n,f)-Z_0}{Z_n(C_n,f)+Z_0}
\end{equation}
where $Z_0 = 377\,\Omega$ is the impedance of free space.
Hence, if a sinusoidal signal with frequency $f$ reaches the IRS element, it will be scattered with a relative amplitude change of $|\Gamma(C_n,f)|$ and phase-shift of $\arg(\Gamma(C_n,f) )$.

\begin{figure}[t!]
        \centering
        \begin{subfigure}[b]{\columnwidth} \centering  
	\begin{overpic}[width=\columnwidth,tics=10]{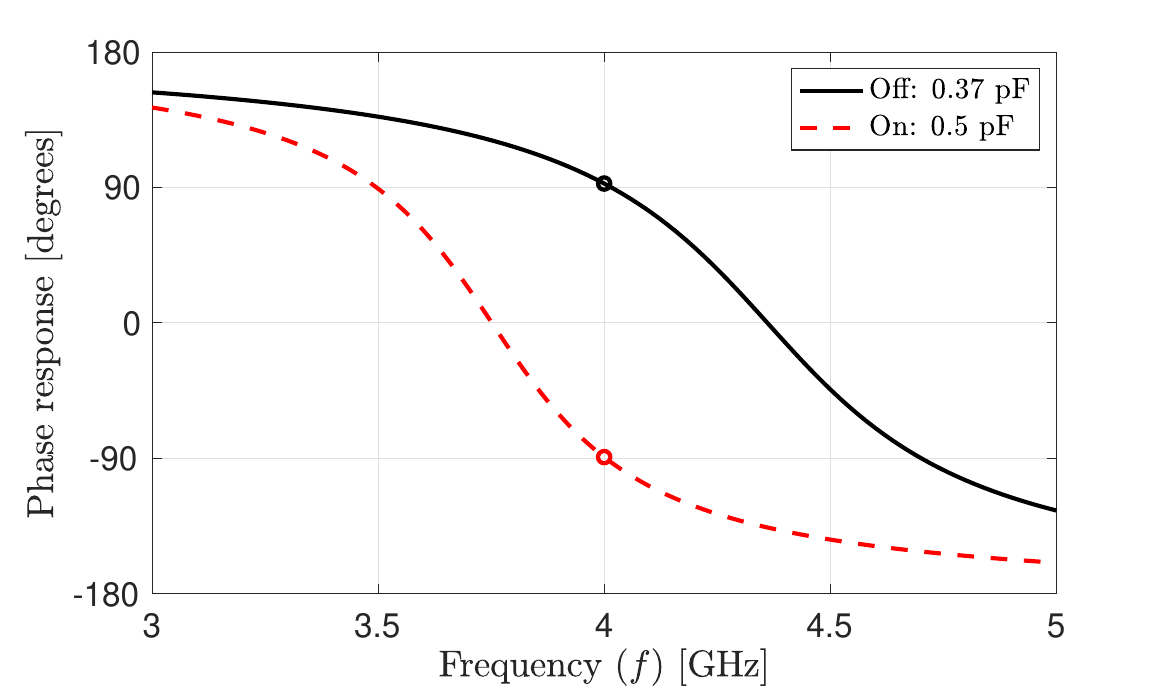}
\end{overpic} 
                \caption{Phase response.}  \vspace{1mm}
                \label{fig:circuit-example:phase}
        \end{subfigure}  
        \begin{subfigure}[b]{\columnwidth} \centering 
	\begin{overpic}[width=\columnwidth,tics=10]{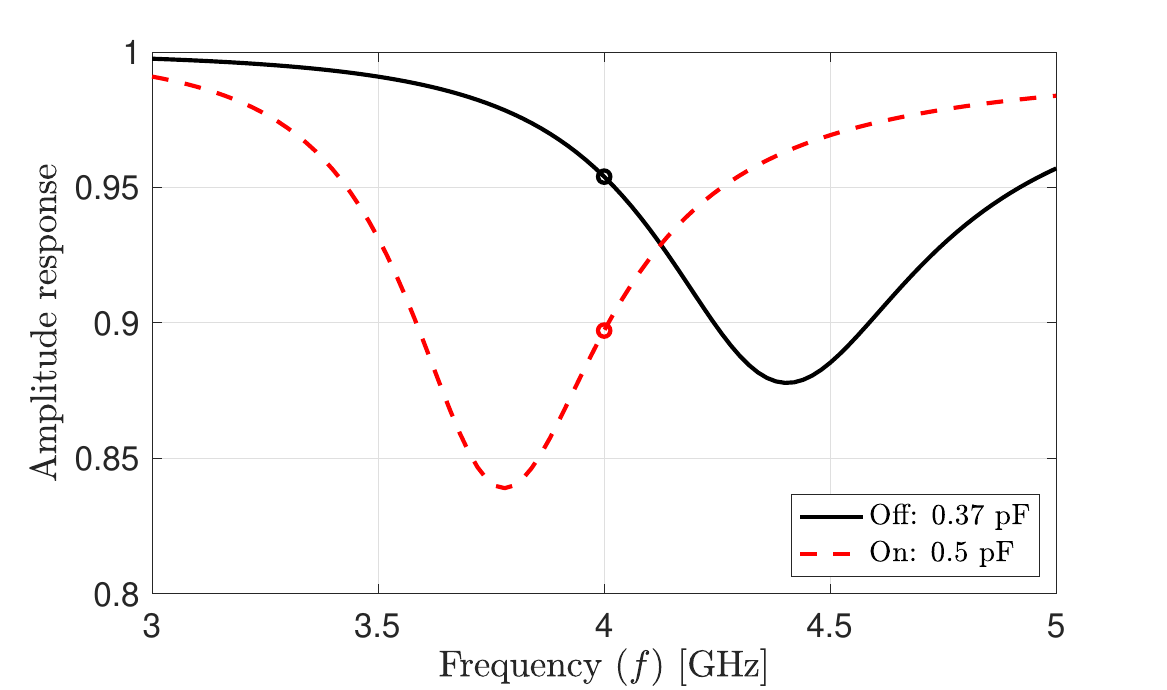}
\end{overpic}
                \caption{Amplitude response.} 
                \label{fig:circuit-example:amplitude}
        \end{subfigure}
        \caption{The reflection coefficient in \eqref{eq:reflection-coefficient} is different for the on/off states, resulting in phase-shifts at the $4$\,GHz frequency that are roughly $180^\circ$ apart, but also unbalanced amplitudes.} \vspace{-3mm}
        \label{fig:circuit-example}
\end{figure}

The IRS is controlled by a diode that can assign two different capacitances: $C_{\textrm{off}}= 0.37$\,pF and $C_{\textrm{on}}=0.5$\,pF.
Fig.~\ref{fig:circuit-example} shows the amplitude and phase of the reflection coefficient, for a range of frequencies around the carrier. By design, the two states are  providing phase shifts of approximately $+90^\circ$ and  $-90^\circ$, respectively. Hence, each element can shift the phase of the reflected signal by $180^\circ$ by turning on the diode. 
The amplitude is unbalanced between the two states, as is typically the case in practice, with the on-state providing a 10\% amplitude loss and the off-state providing a 5\% loss.

\begin{figure}[t!]
	\centering 
	\begin{overpic}[width=\columnwidth,tics=10]{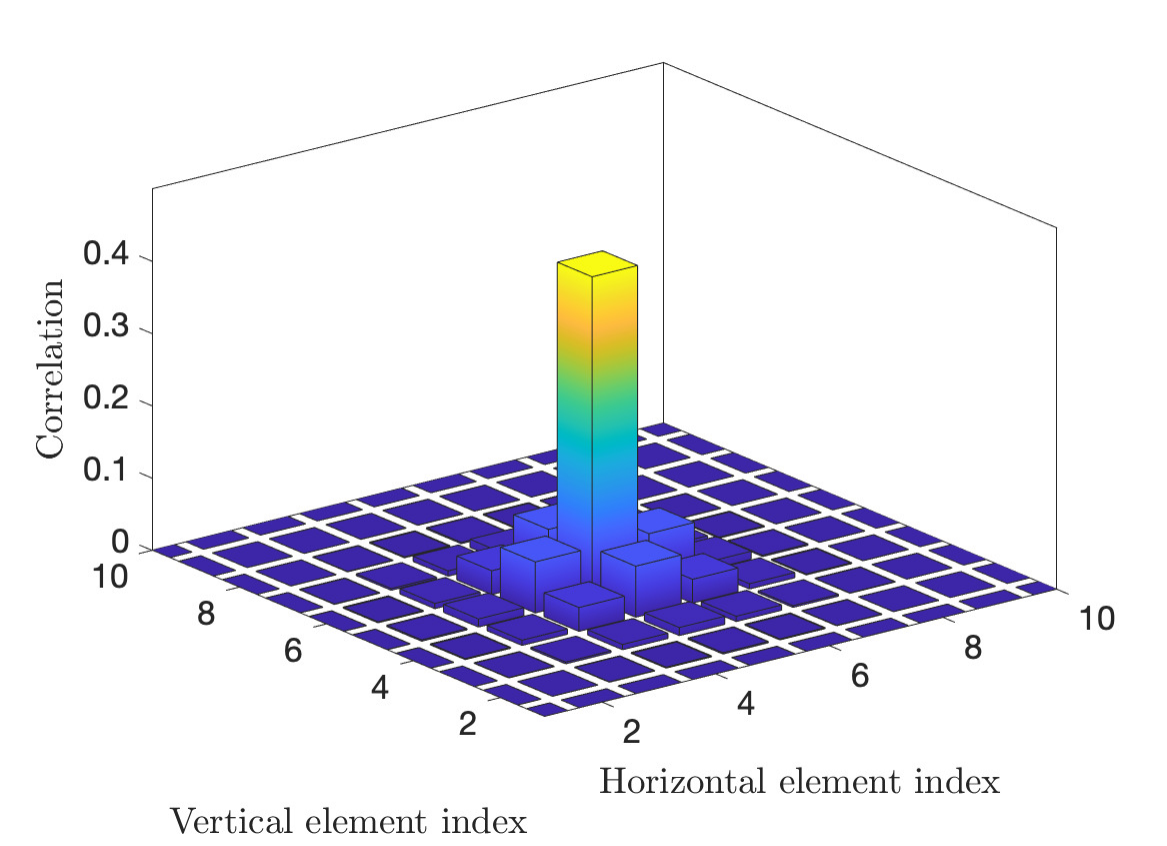}
\end{overpic} 
	\caption{The actual capacitance of the IRS element at row 5 and column 5 is correlated with the intended capacitances of itself and the neighboring elements, according to the pattern illustrated in this figure and specified in \eqref{C_n_correlation}.}
	\label{figure_correlation_function} 
\end{figure}

Another practical complication is the coupling between adjacent IRS elements. It is modeled as a leakage that makes the capacitance $C_n$ of a particular IRS element a function of not only the local state, but also the states of the surrounding elements. 
Let $\widetilde{C}_{n,\vect{\theta}} \in \{ C_{\textrm{off}},C_{\textrm{on}}\}$ be the intended capacitance of element $n$. The actual capacitance is then computed as\footnote{This is a statistical correlation model of the coupling effect, not a physically exact model since such are yet to be developed for IRS \cite{RIS_SPMAG}.}
\begin{equation} \label{C_n_correlation}
C_n = \sum_{i=1}^{N} \widetilde{C}_{i,\vect{\theta}}  \frac{100^{-d_{n,i}/\lambda}}{\sum_{j=1}^{N} 100^{-d_{n,j}/\lambda}}
\end{equation}
where $d_{n,i} = \| \vect{u}_n - \vect{u}_i \|$ is the distance between element $n$ and element $i$. This coupling structure is illustrated in Fig.~\ref{figure_correlation_function} by considering the element at row $5$ in column $5$. The figure shows the ratio that is multiplied with $\widetilde{C}_{i,\vect{\theta}}$ in \eqref{C_n_correlation} for the different elements. We notice that the actual capacitance is determined to 45\% by its intended capacitance and by 55\% of the intended capacitance of the neighboring elements. The closer two elements are, the more they influence each other.
With this IRS modeling, the reflection coefficients become
\begin{equation}
\vect{\omega}_{\vect{\theta}} = \begin{bmatrix} \Gamma(C_1,f_c) \\ \vdots \\ \Gamma(C_N,f_c)  \end{bmatrix}.
\end{equation}

Even if there are only two states per element, the total number of IRS configurations is $2^N =2^{4096} \approx 10^{1233}$, which is practically infinite (e.g., greater than the number of atoms in the observable universe). Hence, finding a rate-maximizing IRS configuration by a global optimization method is hopeless. To develop a heuristic solution, we need to make use of the underlying properties of the propagation environment.

\subsection{Propagation Environment}

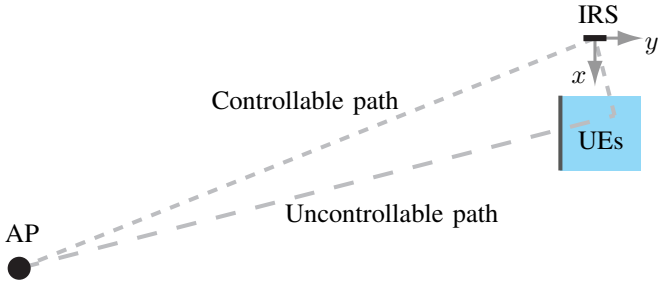
\begin{figure}[t!]
	\centering 
	\begin{overpic}[width=\columnwidth,tics=10]{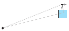}
		\put(88,41){IRS}
		\put(88,22){UEs}
		\put(87,32){$x$}
		\put(98,37){$y$}
		\put(2,8){AP}
		\put(33,28){Controllable path}
		\put(44,11){Uncontrollable path}
\end{overpic} 
	\caption{The simulation setup considers UEs in an area close to an IRS. The users are served by an AP that is located further away. There is LOS propagation between the AP and IRS and NLOS propagation between the AP and UEs. The channel between the IRS and UEs can be either LOS or NLOS.}
	\label{figure_simulation_geometry}  
\end{figure}

We consider the deployment scenario illustrated in Fig.~\ref{figure_simulation_geometry}. The IRS is positioned in the origin (using the coordinate system defined in Fig.~\ref{figure_geometric_setup}), while the AP is located at $[40,-100,0]^{\Ttran}$ (in meters) and the UEs are within a square of size $13$\,m $\times 14$\,m in the azimuth plane centered around $[16.5,1,0]^{\Ttran}$. The uncontrollable direct path between the AP and any UE is assumed to feature non-line-of-sight (NLOS) propagation (as illustrated by a blocking wall), which is a main motivation for deploying an IRS to aid the communications \cite{RIS_SPMAG}. The IRS is deployed to feature line-of-sight (LOS) propagation from the AP and also to most of the UEs, but we will also consider the case of NLOS propagation, which could happen when the LOS path is blocked by some object close to the UE.

We assume all the scattering objects are also located in the horizontal plane, thus the array response vector in \eqref{array-response} can be simplified by dropping the dependence on the elevation angle:
\begin{equation}\label{array-response2}
\vect{a}(\varphi) = \cos(\varphi) \left[e^{\imagunit 0.8\pi    \sin(\varphi) i(1)},\ldots,e^{\imagunit 0.8\pi   \sin(\varphi) i(N) }\right]^{\Ttran}.
\end{equation}
We can notice that the phase-shifts only depend on the horizontal indices $i(n)$, which means that all elements in the same column of the IRS will receive identical signal copies when a wave impinges from the azimuth angle $\varphi$.

We utilize the wideband channel modeling from \cite{RIS_SPMAG}. More precisely, the uncontrollable direct path is given by
\begin{equation}
\vect{h}_d =\sum_{l=1}^{L_d} \sqrt{\beta_{d,l}} e^{-\imagunit 2\pi f_c \tau_{d,l}} \begin{bmatrix} 
\sinc ( 0 + B( \eta -\tau_{d,l}) ) \\ 
\vdots \\
\sinc ( M-1 + B( \eta -\tau_{d,l}) )
\end{bmatrix},
\end{equation}
where $L_d$ is the number of propagation paths, $\beta_{d,l} \geq 0$ is the pathloss of the $l$th path, and $\tau_{d,l}$ is the propagation delay.
The parameter $\eta$ is the sampling delay at the receiver side, which is selected as $\eta = \min_{l} \tau_{d,l}$ to take the first sample when the signal arriving over the shortest path.

Similarly, the controllable path is represented by the matrix
\begin{align}  \notag
\vect{V} = \sum\limits_{l=1}^{L_a}  \sum\limits_{\ell=1}^{L_b} &
\sqrt{\beta_{a,l} \beta_{b,\ell} } e^{-\imagunit 2\pi f_c (\tau_{a,l}+\tau_{b,\ell})}
(\vect{a}(\varphi_{a,l}) \odot \vect{a}(\varphi_{b,\ell})) \\ &\times \begin{bmatrix}
\sinc \big( 0 + B( \eta -\tau_{a,l}-\tau_{b,\ell}) \big) \\
\vdots \\
\sinc \big( M-1 + B( \eta -\tau_{a,l}-\tau_{b,\ell} \big) \\
\end{bmatrix}^{\Ttran},
\end{align}
where there are $L_a$ propagation paths from the AP to the IRS and $L_b$ propagation paths from the IRS to the UE.
The $l$th path from the AP to the IRS is represented by the pathloss $\beta_{a,l} \geq 0$, the propagation delay $\tau_{a,l}$ to the IRS element in the origin, and the incident angle $\varphi_{a,l}$ that determines the phase-shifts at the remaining IRS elements (based on the array response vector $(\vect{a}(\varphi_{a,l})$).
The $\ell$th path from the IRS to the UE is represented by the pathloss $\beta_{b,\ell} \geq 0$, the propagation delay $\tau_{b,\ell}$ from the IRS element in the origin, and the angle-of-departure $\varphi_{b,\ell}$ that determines phase-shifts at the remaining IRS elements.

The parameter values are generated by adapting the 3GPP channel model in \cite{3GPP25996} to scenario at hand and the exact details are provided in the accompanying simulation code \cite{SP_CUP2021}. There are $K = 500$ subcarriers and the time-domain FIR filter has $M = 20$ taps. We  consider 51 UEs, whereof 14 are in NLOS.

\subsection{Preliminaries on Channel Estimation}

Pilots are transmitted to enable optimized IRS configuration.
During the pilot transmission, all the elements of the  signal $\vect{\bar{x}}$ in \eqref{system-model} are $\sqrt{P/B}$, where $P/B$ is the power per transmitted symbol computed based on the transmit power $P=1$\,W and the symbol rate $B=10^7$. The received signal when using the IRS configuration $\vect{\theta}_c$ is then obtained from \eqref{system-model} as
\begin{equation} \label{system-model-pilot}
\vect{\bar{z}}_c = \sqrt{\frac{P}{B}} \vect{\bar{h}}_{\vect{\theta}_c}  + \vect{\bar{w}}_c,
\end{equation}
where the subscript $c$ has been added to indicate which configuration is considered.
From pilot transmission $c$, we can directly compute a least-squares (LS) estimate of the resulting channel $\vect{\bar{h}}_{\vect{\theta}_c}$ with configuration $\vect{\theta}_c$ as 
\begin{equation} \label{eq:LS-estimate-effective-channel}
\hat{\vect{h}}_{\vect{\theta}_c} =  \sqrt{\frac{B}{P}} \vect{\bar{z}}_c  = \vect{\bar{h}}_{\vect{\theta}_c}  + \sqrt{\frac{B}{P}} \vect{\bar{w}}_c.
\end{equation}
To identify a preferable IRS configuration, we could transmit pilots with $C$ different configurations $\vect{\theta}_1,\ldots,\vect{\theta}_C$, estimate the resulting channels using \eqref{eq:LS-estimate-effective-channel}, and then pick the one that provides the highest rate according to \eqref{eq:rate-OFDM}. However, since there are $2^N$ possible configurations, $C$ would have to be huge if this approach should have a reasonable chance of finding a configuration that is close to the optimal one.
A more effective strategy is to utilize the pilot transmission to estimate the underlying channel components $\vect{h}_d$ and $\vect{V}$. 
Since each IRS element can only take two states, the pilot signaling will be based on utilizing the columns of an $N \times N$ Hadamard matrix $\vect{H}_N$, whose entries are either $+1$ or $-1$ with a pattern that makes the columns mutually orthogonal; that is, $\vect{H}_N^{\Htran} \vect{H}_N = N \vect{I}_N$. We notice that these two entries are separated by $180^\circ$ in the complex domain, just as the two IRS states illustrated in Fig.~\ref{fig:circuit-example}, which makes it possible to transmit them by mapping $+1$ to ``on'' and $-1$ to ``off'' (or the other way around).\footnote{Strictly speaking, we consider the pilot matrix is $e^{-\imagunit \pi/2}\vect{H}_N$ but any scalar constant can be absorbed into $\vect{V}$ without loss of generality.} We will use this matrix for pilots in the sequel.

By utilizing the channel structure in \eqref{eq:OFDM-response}, the received signal from the pilot transmission can be expressed as
\begin{align} \notag
\vect{Z} &= \sqrt{\frac{P}{B}}  \vect{F} \left( \vect{h}_d [1,\ldots,1] + \vect{V}^{\Ttran} [\vect{\omega}_{\vect{\theta}_1},\ldots,\vect{\omega}_{\vect{\theta}_C}] \right)  + \vect{W} \\  \label{eq:received-pilot}
&= \sqrt{\frac{P}{B}}   \vect{F} \begin{bmatrix} \vect{h}_d,  \vect{V}^{\Ttran} \end{bmatrix} \begin{bmatrix} 1, \ldots,  1 \\ \vect{\Omega}\end{bmatrix} + \vect{W} 
\end{align}
where $\vect{Z} = [\vect{\bar{z}}_1,\ldots,\vect{\bar{z}}_C] \in \mathbb{C}^{N \times C}$ contains all the received signals, $\vect{W} = [\vect{\bar{w}}_1,\ldots,\vect{\bar{w}}_C] \in \mathbb{C}^{N \times C}$ is the noise matrix, and $\vect{\Omega} = [\vect{\omega}_{\vect{\theta}_1} , \ldots , \vect{\omega}_{\vect{\theta}_C}]$ gathers all the IRS configurations.
Ideally, $\vect{\Omega}$ is a known matrix but this is not the case in practice due to coupling and other hardware effects.
Although $\vect{\theta}_1, \ldots, \vect{\theta}_C$ are known, the mapping to $\vect{\Omega}$ is only approximately known.

We  express the mismatch as \vspace{-1mm}
\begin{equation}
\vect{\Omega} = \hat{\vect{\Omega}} + \vect{E},
\end{equation}
where is $\hat{\vect{\Omega}} $ is the intended matrix and $ \vect{E} = \vect{\Omega} - \hat{\vect{\Omega}}$ is the unknown mismatch.
Let $^{\dagger}$ denote the Moore-Penrose inverse.
An LS estimate of $\vect{h}_d$ and $\vect{V}$ can be obtained from \eqref{eq:received-pilot} as  \vspace{-2mm}
\begin{align} \notag
&\sqrt{\frac{B}{P}} \vect{F}^{\dagger} \vect{Z} \begin{bmatrix} 1, \ldots,  1 \\ \hat{\vect{\Omega}} \end{bmatrix} ^{\dagger} 
= \underbrace{\begin{bmatrix} \vect{h}_d,  \vect{V}^{\Ttran} \end{bmatrix}}_{\textrm{Desired channel}} \underbrace{\begin{bmatrix} 1, \ldots,  1 \\ \hat{\vect{\Omega}} \end{bmatrix}}_{=\hat{\vect{\Omega}}_e} \underbrace{\begin{bmatrix} 1, \ldots,  1 \\ \hat{\vect{\Omega}}\end{bmatrix}^{\dagger}}_{=\hat{\vect{\Omega}}_e^{\dagger}} \\
&\!+\!  \underbrace{\begin{bmatrix} \vect{h}_d, \vect{V}^{\Ttran} \end{bmatrix} \begin{bmatrix} 0, \ldots,  0 \\ \vect{E}\end{bmatrix} \begin{bmatrix} 1, \ldots,  1 \\ \hat{\vect{\Omega}}\end{bmatrix}^{\dagger}}_{\textrm{Hardware mismatch}}  \! + \underbrace{\sqrt{\frac{B}{P}} \vect{F}^{\dagger} \vect{W} \begin{bmatrix} 1, \ldots,  1 \\ \hat{\vect{\Omega}} \end{bmatrix} ^{\dagger}}_{\textrm{Noise}} \! .\label{eq:estimate-LS}
\end{align} \vskip-3mm
\noindent It is only if the extended pilot matrix $\hat{\vect{\Omega}}_e$ has full rank that $\hat{\vect{\Omega}}_e \hat{\vect{\Omega}}_e^{\dagger} = \vect{I}_{N+1}$ so that we obtain an (over)determined system with a unique solution.\footnote{The matrix $\vect{F}$ must also have full rank but it is satisfied by design.}
In Section~\ref{sec:IRS-configuration}, we will show how to handle both situations in the considered system.

\subsection{Dataset}

We will demonstrate how one can estimate and configure an IRS under the conditions described above, without having explicit prior knowledge about the setup. 
This was the task of the IEEE Signal Processing Cup 2021 \cite{SP_CUP2021}, where the competing teams were given a dataset with received signals from the pilot transmission. 
There are two datasets: one with extensive pilot signaling for a single UE and the other with shorter pilot signaling for 50 UEs. The dataset and code is available at \url{https://github.com/emilbjornson/SP_Cup_2021}

\begin{enumerate}
\item \textbf{Dataset1:} The dataset contains $C=4N$ received OFDM signal blocks of the kind in \eqref{system-model-pilot} obtained with the pilot matrix $\hat{\vect{\Omega}} = [\vect{H}_N, - \vect{H}_N, \vect{H}_{N,\textrm{flip}}, - \vect{H}_{N,\textrm{flip}}]$, where we recall that $\vect{H}_N$ denotes the $N \times N$ Hadamard matrix and $\vect{H}_{N,\textrm{flip}}$ denotes the resulting matrix when each column is flipped upside-down. This represents the signals received at a single UE, to enable estimation of both the channel and other parameters.

\item \textbf{Dataset2:} The dataset considers 50  UEs and contains $C=N$ received OFDM signal blocks per UE of the kind in \eqref{system-model-pilot} obtained with the pilot matrix $\hat{\vect{\Omega}}=\vect{H}_N$.

\end{enumerate}

\section{Novel IRS Configuration Methods}
\label{sec:IRS-configuration}

In this section, we will provide novel methods for IRS configuration that can be utilized when having two IRS states with unbalanced and unknown amplitudes.

\subsection{Estimation of the Noise Statistics}

The power spectral density $N_0$ of the noise is often treated as given but must be estimated in practice. In theory, one can let the transmitter be silent and then measure the received signal, which should only contain noise. However, this might not result in an accurate noise estimate since the transceiver hardware acts differently depending on the dynamic range of the received signal.
Instead, we will consider the difference between received signals. 
With mutual coupling, one can only be sure to fully remove the signal components if the IRS configurations are identical.
In Dataset1, there are many IRS configurations that are utilized multiple times, thus one can compute the difference and divide by two to get a variable that only contains the noise.
By following this approach, we can easily obtain an estimate of $N_0$ that is very close to the true value that was used to generate the datasets:
\begin{equation}
N_0 = 10^{\frac{-204+9}{10}}\, \, \textrm{W/Hz}
\end{equation}
which in this setup is 9\,dB larger than the thermal noise floor at room temperature. This is the noise figure in the UE hardware.

\subsection{Discovering the Spatial Structure}

\begin{figure}[t!]
	\centering 
	\begin{overpic}[width=.85\columnwidth,tics=10]{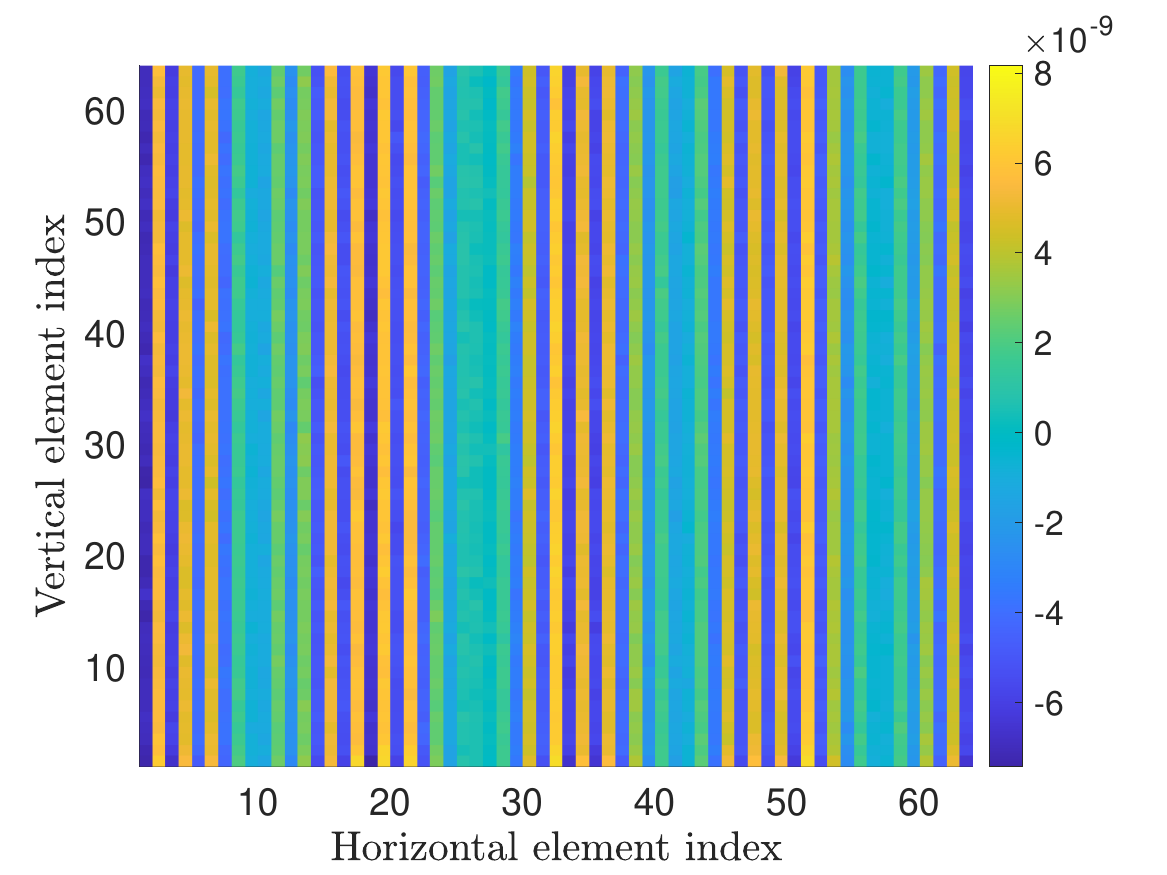}
\end{overpic} 
	\caption{The real part of the estimates of the first column in $\vect{V}$ from Dataset1 reveals that the channel coefficients are approximately equal within each column of the array.} \vspace{-3mm}
	\label{figure_realparts}  
\end{figure}

The number of pilots in Dataset1 is sufficient to apply the LS estimate in \eqref{eq:estimate-LS} to uniquely estimate $\vect{h}_d$ and $\vect{V}$, because $C = 4N > N+1$ and thus $\hat{\vect{\Omega}}_e$ has full rank.
The estimate can be utilized to discover the spatial structure of the array and propagation environment. Fig.~\ref{figure_realparts} shows the real part of the estimates of the first column in $\vect{V}$ on a two-dimensional grid, which reveals that the estimates are nearly the same within each column and widely different between columns. This structure can be utilized to identify both the shape of the array (if it was unknown) and spatial correlation that can be utilized to reduce the channel dimension and estimation overhead. In this case, we notice that the channel coefficient is the same within each column, which is in line with the channel modeling but in practice, any spatial structure can be identified in this way and utilized to simplify the estimation.

\subsection{Dimension Reduction}

The number of pilots in Dataset2 is insufficient to obtain a unique LS estimate since  $\hat{\vect{\Omega}}_e$ cannot have full rank when $C = N < N+1$.
However, we can reduce the dimension of the estimation problem by exploiting the fact that the channel coefficients within each column of the IRS are equal. We can rewrite $\vect{V}$ as
\begin{equation} \label{eq:subspace-reduction}
\vect{V} = \underbrace{(\vect{1}_{N_{\mathrm{V}}} \kron \vect{I}_{N_{\mathrm{H}}})}_{=\vect{A}} \vect{V}_{\textrm{row}}
\end{equation}
where $\vect{A}=(\vect{1}_{N_{\mathrm{V}}} \kron \vect{I}_{N_{\mathrm{H}}}) \in \mathbb{C}^{N \times N_{\mathrm{H}}}$, $\kron$ denotes the Kronecker product, and $\vect{1}_{N_{\mathrm{H}}} = [1,\ldots,1]^{\Ttran} \in \mathbb{C}^{N_{\mathrm{H}}}$. Most importantly, $\vect{V}_{\textrm{row}} \in \mathbb{C}^{N_{\mathrm{H}} \times M}$ is the reduced-dimension matrix that captures the channel coefficients for any row of the IRS.
By substituting \eqref{eq:subspace-reduction} into \eqref{eq:received-pilot}, we obtain
\begin{align}  \label{eq:received-pilot-reduced}
\vect{Z}= \sqrt{\frac{P}{B}}   \vect{F} \begin{bmatrix} \vect{h}_d, \vect{V}_{\textrm{row}}^{\Ttran} \end{bmatrix} \begin{bmatrix} 1, \ldots,  1 \\  \vect{A}^{\Ttran} \vect{\Omega} \end{bmatrix} + \vect{W}.
\end{align}
The reduced dimension makes it possible to follow the procedure  in \eqref{eq:estimate-LS}
 to compute the unique LS estimates $\hat{\vect{h}}_d$ and $\hat{\vect{V}}_{\textrm{row}}$ as 
\begin{equation}
 \begin{bmatrix} \hat{\vect{h}}_d, \hat{\vect{V}}_{\textrm{row}}^{\Ttran} \end{bmatrix}  = 
\sqrt{\frac{B}{P}} \vect{F}^{\dagger} \vect{Z} \begin{bmatrix} 1, \ldots,  1 \\ \vect{A}^{\Ttran}\hat{\vect{\Omega}} \end{bmatrix} ^{\dagger}.
\end{equation}

\subsection{Optimized Binary IRS Configuration in OFDM Systems}

Finding a suitable IRS configuration with affordable complexity is challenging in OFDM systems due to the tradeoff between the many subcarriers. The strongest tap maximization method is considered in \cite{Zheng2020,Lin2020a}, with the goal of maximizing the magnitude of the largest entry of $\vect{h}_d + \vect{V}^{\Ttran} \vect{\omega}_{\vect{\theta}}$. This works well for LOS channels but not for NLOS channels \cite{RIS_SPMAG}, and the extension to the case with only two states per IRS element is non-trivial. 
We will instead target to maximize the total received signal power, which is proportional to
\begin{align} \notag
\| \bar{\vect{h}}_{\vect{\theta}} \|^2 & = \! \begin{bmatrix} 1 \\ \frac{\vect{A}^{\Ttran} \vect{\omega}_{\vect{\theta}} }{N_{\mathrm{V}}} \end{bmatrix}^{\Htran} \! \begin{bmatrix} \vect{h}_d, N_{\mathrm{V}} \vect{V}_{\textrm{row}}^{\Ttran} \end{bmatrix}^{\Htran} \begin{bmatrix} \vect{h}_d, N_{\mathrm{V}} \vect{V}_{\textrm{row}}^{\Ttran} \end{bmatrix} \! \begin{bmatrix} 1 \\ \frac{\vect{A}^{\Ttran} \vect{\omega}_{\vect{\theta}}}{N_{\mathrm{V}}}  \end{bmatrix} \\
&= \vect{c}^{\Htran} \underbrace{\begin{bmatrix} \hat{\vect{h}}_d, N_{\mathrm{V}} \hat{\vect{V}}_{\textrm{row}}^{\Ttran} \end{bmatrix}^{\Htran} \begin{bmatrix} \hat{\vect{h}}_d, N_{\mathrm{V}} \hat{\vect{V}}_{\textrm{row}}^{\Ttran} \end{bmatrix}}_{=\vect{B}} \vect{c} + \textrm{error},
\end{align}
where the error term contains everything that is unknown.
The known part $\vect{c}^{\Htran} \vect{B} \vect{c}$ is a quadratic form that is to be maximized.
If there would only be a norm constraint, the optimal $\vect{c}$ is the dominant eigenvector of $\vect{B}$. However, that vector will most likely not have binary entries ($\pm 1$) as is required by the considered IRS.
However, we can design a heuristic algorithm by taking inspiration by the \emph{power method} that find the dominant eigenvalue by iterating the computation $\vect{c}_{i+1} = \frac{\vect{B} \vect{c}_i}{\| \vect{B} \vect{c}_i \|}$ from some initial vector $ \vect{c}_0$ until convergence.

The proposed algorithm is to start from some initial solution $\vect{c}_0$ (e.g., all states are equal) and then:
\begin{enumerate}
\item Compute $\vect{d}_{i} = \vect{B} \vect{c}_i$ and let $d_{i,1}$ denote the first entry.
\item Set $\vect{c}_{i+1} = \mathrm{sign}( d_{i,1}^* \vect{d}_{i} )$, where the function outputs $+1$ if the real part is positive and $-1$ otherwise.
\item Iterate until convergence.
\end{enumerate}
This is essentially the power method with an extra step that project the solution onto the feasible set of IRS configurations. The multiplication with $d_{i,1}^*$ in Step 2 ensures that the uncontrollable path is always assigned $+1$.

\begin{figure}[t!]
	\centering 
	\begin{overpic}[width=\columnwidth,tics=10]{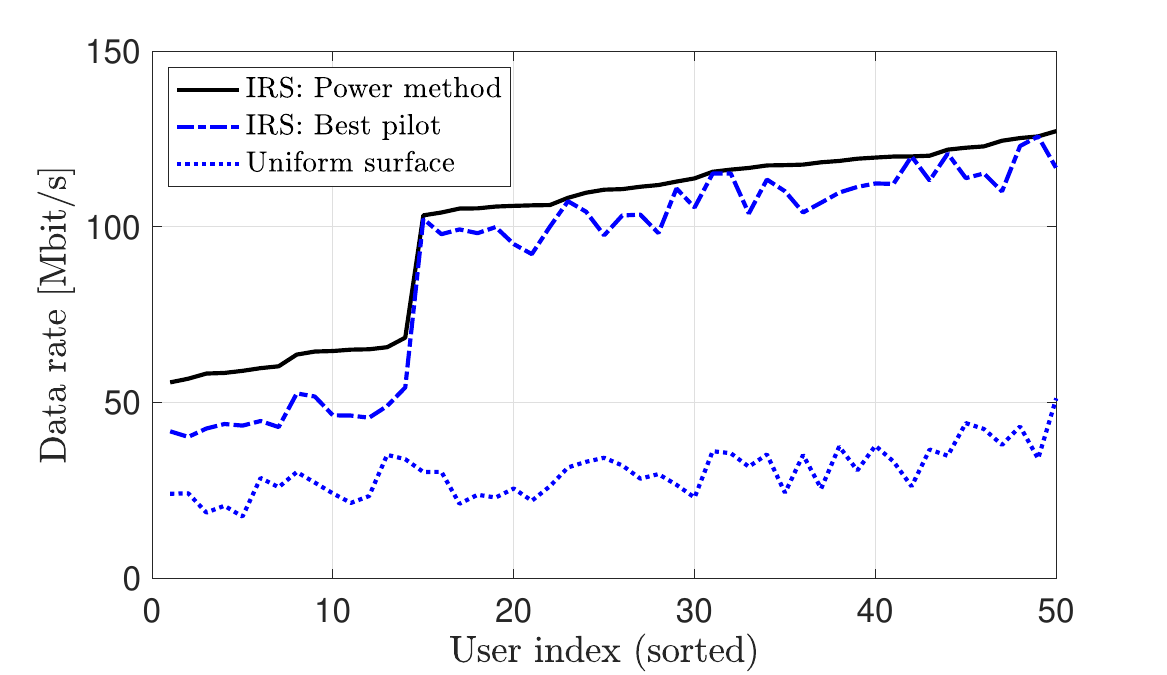}
		\put(15,29){NLOS}
		\put(47,44){LOS}
\end{overpic} 
	\caption{The data rates achieved when using the dataset in \cite{SP_CUP2021}.}
	\label{figure_user_rates}  
\end{figure}

Fig.~\ref{figure_user_rates} shows the rates achieved by the 50 UEs in Dataset2, ordered based on increasing rates with the proposed method. Two benchmarks are considered: 1) select the best configuration among those $N$ considered during pilot transmission; 2) set all coefficients to ``off'' to mimic a uniform metal surface.
The proposed method provides larger rates for all UEs, particularly in NLOS.
The sum rate is $3.3\times$ larger with an optimized IRS compared a uniform surface, which shows the great benefits of the technology even in complicated setups.

\section{Conclusion}

This paper has demonstrated how an IRS can be configured to give decisive performance gains over passive surfaces in practical scenarios with wideband channels, mutual coupling, and binary phase-shifts with unbalanced amplitudes. New channel estimation and configuration algorithms have been proposed and analyzed using the dataset from \cite{SP_CUP2021}. The code is available at \url{https://github.com/emilbjornson/SP_Cup_2021}

\bibliographystyle{IEEEtran}
\bibliography{IEEEabrv,refs}

\end{document}